\documentclass[10pt, twocolumn, superscriptaddress, amsmath, amssymb, aps, longbibliography, pra]{revtex4-2}

\usepackage{graphicx}
\usepackage[caption=false]{subfig}
\usepackage{graphics}
\usepackage{adjustbox} 
\usepackage{dcolumn}
\usepackage{bm}
\usepackage{dsfont}
\usepackage{amsthm}
\usepackage[dvipsnames]{xcolor}
\usepackage{chngcntr}
\usepackage{comment}
\usepackage{apptools}
\AtAppendix{\counterwithin{lemma}{section}}
\usepackage{xy}

\usepackage{mathtools}
\usepackage{floatrow} 
\DeclareFloatFont{tiny}{\tiny}
\usepackage {hyperref}

\usepackage[capitalise]{cleveref}
\usepackage[T1]{fontenc}
\crefname{figure}{Fig.}{Fig.}

\theoremstyle{definition}

\theoremstyle{definition}

\newcommand{\be}{\begin{equation}}
\newcommand{\ee}{\end{equation}}
\newcommand{\bp}{\begin{pmatrix}}
\newcommand{\ep}{\end{pmatrix}}
\newcommand{\ben}{\begin{enumerate}}
\newcommand{\een}{\end{enumerate}}

\begin{document}

\title{Solovay Kitaev Algorithm and Randomized Compilation}

\author{Oliver Maupin}
\affiliation{Quantum Performance Laboratory,
            Sandia National Laboratories,
            Albuquerque, NM, 87185, and
            Livermore, CA, 94550, USA}

\author{Ashlyn D. \surname{Burch}}
\thanks{Present address: Oak Ridge National Laboratory,
            Oak Ridge, TN 37830, USA}
\affiliation{Sandia National Laboratories,
            Albuquerque, NM, 87185, USA}

\author{Christopher G. \surname{Yale}}
\affiliation{Sandia National Laboratories,
            Albuquerque, NM, 87185, USA}
             
\author{Matthew N. H. Chow}
\thanks{Present address: HRL}
\affiliation{Sandia National Laboratories,
             Albuquerque, NM, 87185, USA}
\affiliation{Department of Physics and Astronomy,
             University of New Mexico,
             Albuquerque, NM, 87131, USA}
             
\author{Terra Colvin, Jr.}
\affiliation{Department of Physics and Astronomy, 
             Tufts University, 
             Medford, MA, 02155, USA}

\author{Brandon Ruzic}
\affiliation{Sandia National Laboratories,
             Albuquerque, NM, 87185, USA}

\author{Melissa C. \surname{Revelle}}
\affiliation{Sandia National Laboratories,
            Albuquerque, NM, 87185, USA}

\author{Brian K. \surname{McFarland}}
\affiliation{Sandia National Laboratories,
            Albuquerque, NM, 87185, USA}

\author{Eduardo \surname{Ibarra-Garc\'ia-Padilla }}
\affiliation{Sandia National Laboratories,
            Albuquerque, NM, 87185, USA}

\author{Alejandro \surname{Rascon}}
\affiliation{Sandia National Laboratories,
             Albuquerque, NM, 87185, USA}
 \affiliation{Department of Physics and Astronomy,
             University of New Mexico,
             Albuquerque, NM, 87131, USA}

\author{Andrew J. \surname{Landahl}}
\affiliation{Sandia National Laboratories,
             Albuquerque, NM, 87185, USA}
 \affiliation{Department of Physics and Astronomy,
             University of New Mexico,
             Albuquerque, NM, 87131, USA}

\author{Susan M. \surname{Clark}}
\affiliation{Sandia National Laboratories,
             Albuquerque, NM, 87185, USA}


\author{Peter J. Love}
\affiliation{Department of Physics and Astronomy, 
             Tufts University, 
             Medford, MA, 02155, USA}
\affiliation{Computational Science Initiative,
             Brookhaven National Laboratory,
             Upton, NY, 11973, USA}  

\begin{abstract}
We analyze the use of the Solovay Kitaev (SK) algorithm to generate an ensemble of one qubit rotations over which to perform randomized compilation. We perform simulations to compare the trace distance between the quantum state resulting from an ideal one qubit $R_{Z}$ rotation and discrete SK decompositions. We find that this simple randomized gate synthesis algorithm can reduce the approximation error of these rotations in the absence of gate errors in simulation by at least a factor of two compared to a naive gate synthesis algorithm. We test the technique under the effects of a simple coherent noise model and find that it can mitigate coherent noise. We also run our algorithm on Sandia National Laboratories' QSCOUT trapped-ion device and find that randomization is able to help in the presence of realistic noise sources.
\end{abstract}

\pacs{Valid PACS appear here} 
\maketitle


\section{Introduction}
\label{sec:Introduction}

Quantum computing is currently in the Noisy Intermediate Scale Quantum (NISQ) era of development~\cite{preskill2018quantum}. NISQ computers exhibit quantum coherence, scale to hundreds of qubits, achieve error rates of less than $1\%$, and have enabled many small-scale demonstrations of quantum algorithms, as well as larger scale experiments that push towards utility ~\cite{lau2022nisq}. NISQ machines lack quantum error correction, however a wide range of error mitigation techniques have been developed which aim to extract better results from noisy quantum data~\cite{cai2023quantum}.

To move beyond the NISQ era requires implementation of quantum error correction (QEC)~\cite{lidar2013quantum}. QEC encodes each logical qubit in many physical qubits. If the error rate on the physical qubits is below a threshold (whose value depends on the particular scheme used) then the error rates for all encoded gates on the logical qubits can be lower than those on the physical qubits. This enables continued reduction in error rates at the price of more physical qubits, but without further improvements in physical qubit noise. These techniques are necessary to reach the very large numbers of qubits and gate operations that are required by applications at the quantum advantage scale.

With recent demonstrations of logical qubits in several platforms we are seeing the first demonstrations of post-NISQ quantum technology~\cite{egan2021fault,google2023suppressing,Acharya2025,bluvstein2024logical,pogorelov2025experimental}. This new era of development has been termed Early Fault Tolerant Quantum Computation (EFTQC)~\cite{katabarwa2024early}. Like NISQ, these EFTQC demonstrations will inform the development of future large scale quantum computation. These advancements raise the question: can NISQ experiments inform the development of EFTQC devices? In the present paper we address one aspect of this question by considering a key element of QEC, the decomposition of continuous operations into circuits over a discrete gate.

\begin{figure*}
    \centering
    \subfloat{
        \includegraphics[width=0.49\textwidth]{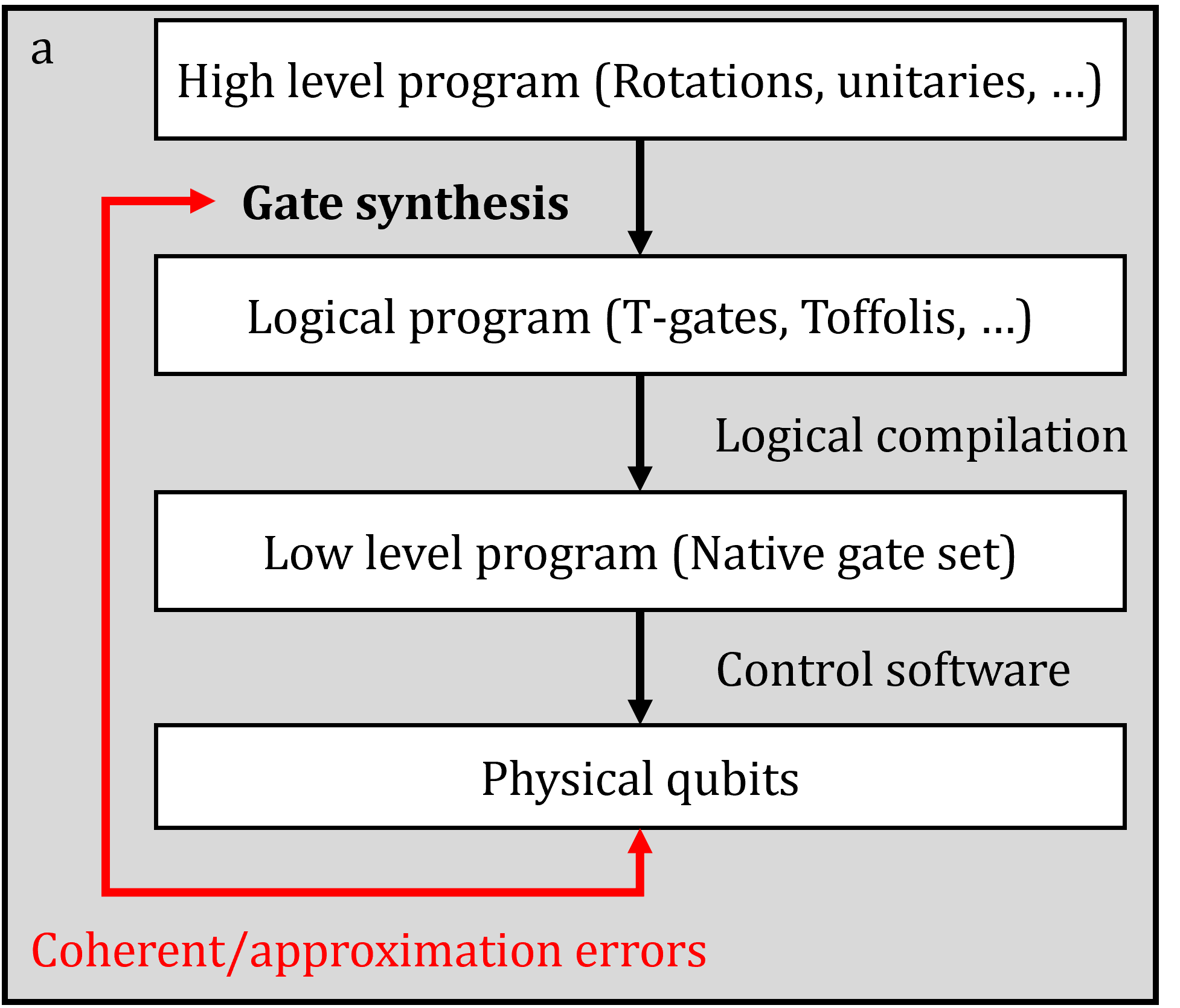}
        \label{sfig:SKARC_QEC_Stack}
    }
    \subfloat{
        \includegraphics[width=0.49\textwidth]{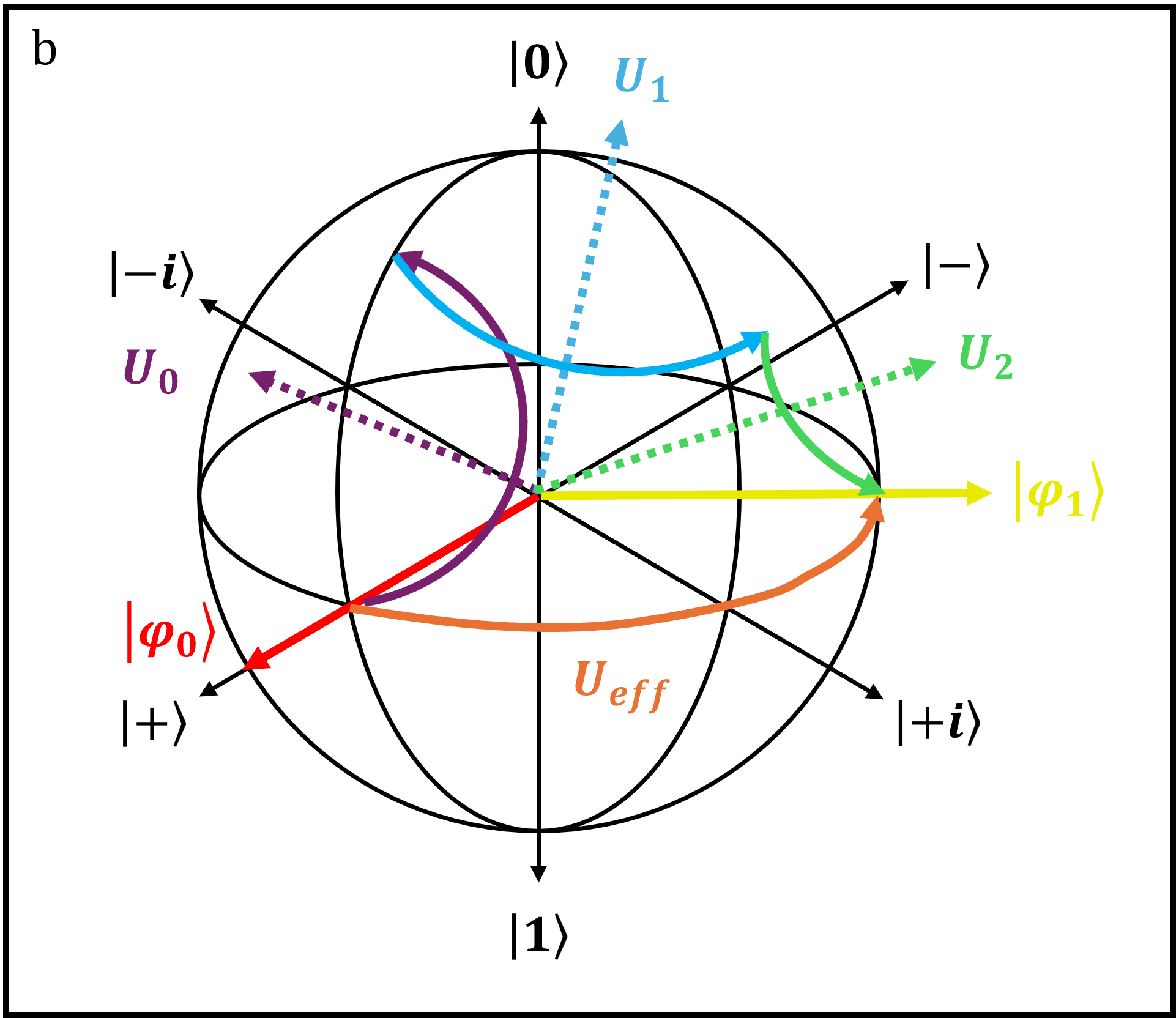}%
        \label{sfig:SKARC_Gate_Synthesis_Diagram}%
    }

    \caption{Overview of gate synthesis. \textbf{(a)} Gate synthesis takes arbitrary operations from a high level program and compiles them into discrete logical operations acting on logical qubits. This work focuses on the effects of coherent errors and approximation errors on gate synthesis and vice-versa. \textbf{(b)} Single qubit gate synthesis decomposes an arbitrary rotation $U_{\textrm{eff}}$ into a series of discrete operations $U_{0}, U_{1}, U_{2}, ... \in \{\mathcal{G} \}$ from a gate-set $\{ \mathcal{G} \}$.}
\end{figure*}

The Solovay-Kitaev (SK) Theorem ~\cite{Kitaev_1997b, Kitaev_1995, Kitaev_2002,Dawson_2006, nielsen_chuang_2010} shows that it is possible to find such decompositions of continuous rotations into a discrete gate set as is required by QEC. This process is called gate synthesis, and is the focus of ongoing research \cite{Koczor_2024, Seiseki_2024, Yoshioka_2025}. This compilation takes place in the quantum computing stack between the high level circuit as outlined by a quantum algorithm, and the logical program expressed in logical operations, as shown in Figures \ref{sfig:SKARC_QEC_Stack} and \ref{sfig:SKARC_Gate_Synthesis_Diagram}. However, this decomposition can be costly, requiring extra circuit depth compared to physical rotations on NISQ devices. When then could an EFTQC outperform arbitrary rotations on a NISQ device?

Current experimental demonstrations of quantum error correction focus on implementing universal and reliable logical operations that outperform physical operations ~\cite{egan2021fault,google2023suppressing,bluvstein2024logical,pogorelov2025experimental}. For these demonstrations to outperform the high fidelities of single qubit gates on NISQ devices, it is important to reduce the cost of compilation, as the required increase in circuit depth penalizes the performance of EFTQC devices ~\cite{egan2021fault}. In this work, we propose a method for decreasing the cost of single qubit gate synthesis in the presence of coherent errors and benchmark its effectiveness using a trapped-ion device as a baseline of performance.

Coherent noise is believed to be corrected or at least mitigated by error correction codes intended for use with Solovay-Kitaev decompositions \cite{Greenbaum_2016, Bravyi_2017}. However, the use of error correcting codes does not rule out coherent errors that can occur in early fault tolerant devices that merely suppress the error rather than eliminating it entirely. This is the case for EFTQC architectures with smaller code sizes \cite{Bravyi_2017}, or architectures wherein cycles of syndrome extraction are costly and may only be applied sparingly throughout the duration of circuits \cite{Berthusen_2025}. In this context, coherent errors at the physical level can manifest as logical errors, impacting performance.

Our method aims to reduce the effects of coherent and approximation errors by using the inherent compilation degeneracy in existing gate synthesis algorithms. Some gate synthesis algorithms attempt to calculate decompositions according to a cost metric, but do not try to perform exact synthesis to find the \textit{most accurate} decomposition \cite{Dawson_2006, Matsumoto_2008, gridsynth, Seiseki_2024}. That is, the user chooses some allowed precision $\epsilon$, and the algorithm finds a decomposition that approximates the desired rotation within that precision, according to some distance measure. These algorithms usually minimize the required number of $T$-gates. This allowed precision $\epsilon$ creates a degeneracy in the choice of decomposition, as there are many sequences $R_{i}$ that achieve the same precision. Imposing a limit on the number of gates or the number of $T$-gates reduces the size of this ensemble of approximate rotations, but still allows for multiple different approximations of the same rotation of the same quality.

This compilation degeneracy allows for the use of NISQ era twirling techniques such as Randomized Compilation (RC), a technique that has been used \cite{Ware_2018, Hashim_2021, Kim_2023} to tailor coherent quantum errors into stochastic errors. The idea of randomized compilation is simple. A logical quantum circuit is duplicated $r$ times, each copy being decomposed into a different random set of operations at the gate level. Given a total sampling budget $N$, each of these circuits are then run on the device $N/r$ times, and the resulting measurement outcomes are averaged together. This procedure maintains the global circuit structure but randomizes the gate sequence to mitigate the effects of coherent errors. Namely, RC via Pauli twirling can tailor coherent error into a Pauli channel, and RC via Clifford twirling can further tailor that error into a purely depolarizing channel.

In much the same way, we can probabilistically sample from an ensemble of approximate rotations generated for each of $k$ arbitrary rotations in our circuit. Choosing one sampled sequence for each rotation produces a single sampled circuit, and enough random instantiations of these sampled circuits will randomize coherent errors when their measurement results are averaged together.

Previous works in gate synthesis \cite{Hastings_2016, Campbell_2017, Seiseki_2024} focus on minimizing approximation error incurred during gate synthesis using probabilistic methods. Probabilistic state synthesis uses a weighted mixture of an ensemble of circuits to generate an approximation to a desired quantum state. The mixture over the ensemble results in better average accuracy than any individual circuit. Probabilistic state synthesis is therefore the state equivalent of randomized compilation (RC) \cite{Wallman_2016}, except the error to be minimized includes the synthesis error and not only the hardware error. This is the same notion that underlies \textit{Multi-Product Formulas} \cite{Zhuk_2024} in Hamiltonian simulation.

In the present paper, we explore the effects of coherent and projection noise in the early fault-tolerant regime in addition to synthesis error when using SK sequences, comparing the accuracy and cost of gate synthesis techniques using Sandia National Laboratories' QSCOUT trapped-ion device. The physical single-qubit gates of trapped-ion systems are significantly better than the two-qubit gates, especially for trapped-ion architectures which display extremely high single-qubit gate fidelities \cite{Clark_2021, Quantinuum_2023, IonQ2024Aria}. This makes them a good baseline for comparison with the low error rates of EFTQC systems, as they are able to achieve long sequences of single qubit gates with high fidelity. This foreshadows future benchmarking efforts on EFTQC devices.


This paper is structured as follows. We begin in Section \ref{sec:Sequence Ensemble} by detailing the algorithm that we use to generate SK sequences and analyzing the resulting ensemble of rotations. We then discuss the coherent noise model we use in Section \ref{sec:Coherent Noise Model}. In Section \ref{sec:Results}, we show results from a noiseless simulation of the technique, simulation under our simple coherent noise model, and experiment on a NISQ trapped-ion quantum computer. Finally, we summarize our findings in Section \ref{sec:Conclusions}.

\begin{figure*}
    \centering
    \includegraphics[width=\textwidth]{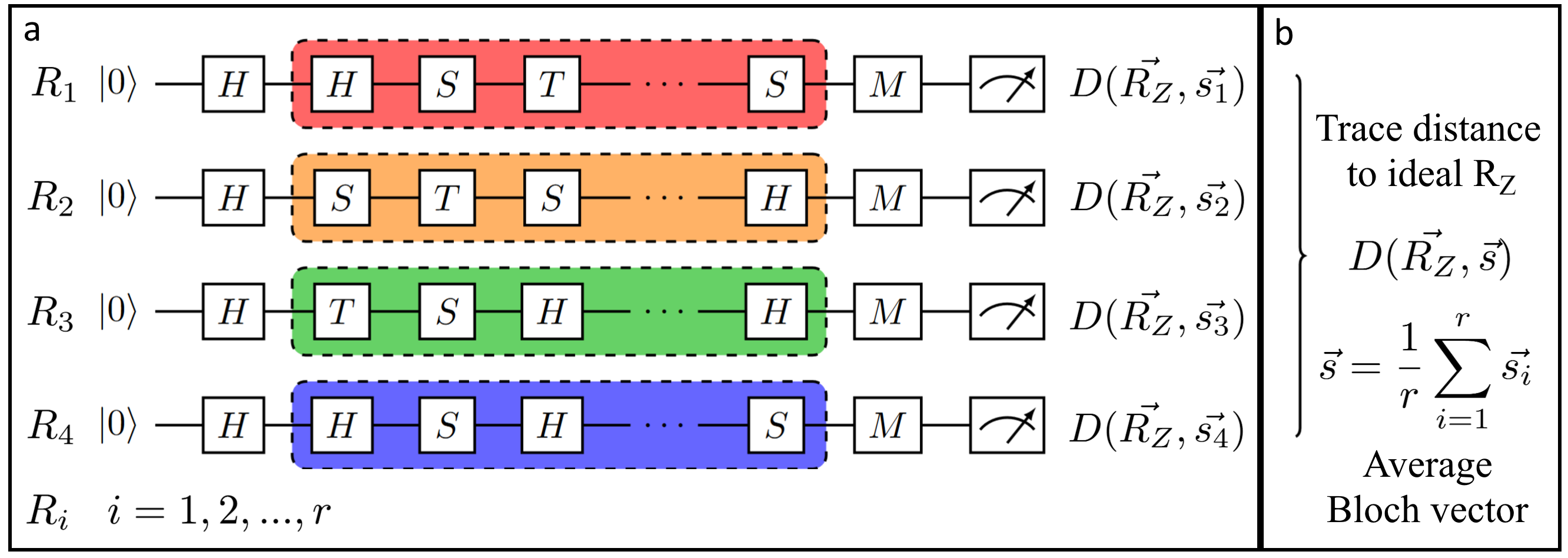}
    \caption{A schematic of our randomization protocol and analysis. \textbf{(a)} Individual rotation sequences $R_{i}$ are generated using gridsynth with slight variations in the initial parameters. The accuracy of each sequence can be evaulated by taking the trace distance the corresponding Bloch vectors $\vec{s_{i}}$ to the exact $R_{Z}$ rotation as measured with state tomography by rotating the state before measurement with Clifford operations $M$. \textbf{(b)} To implement randomization, the measurement results for these sequences are averaged as in Randomized Compiling to achieve an effective rotation and Bloch vector $\vec{s}$. The accuracy of this effective rotation can then be evaluated via trace distance, which is less than the average trace distance for an individual sequence.}
    \label{fig:Randomization Protocol} 
\end{figure*}

\section{Sequence Ensemble}
\label{sec:Sequence Ensemble}

Before we present our results, we first discuss our method for generating distinct SK sequences. We then analyze the characteristics of the resulting ensemble and present our randomization algorithm. The procedure is shown graphically in Figure \ref{fig:Randomization Protocol}.

\subsection{Generating Sequences}
\label{subsec:Generating Sequences}

We consider decompositions into the set of ${ \{H, S, T \}}$ gates, as defined in equation \eqref{eq:HST Gates}:
\begin{equation}
\label{eq:HST Gates}
    H = \frac{1}{\sqrt{2}} \begin{bmatrix}
                                1 & 1 \\
                                1 & -1
                            \end{bmatrix}, \hspace{3mm}
    S = \begin{bmatrix}
        1 & 0 \\
        0 & i
    \end{bmatrix}, \hspace{3mm}
    T = \begin{bmatrix}
        1 & 0 \\
        0 & e^{i \pi/4}
    \end{bmatrix}
\end{equation}
These gates are sufficient to approximate any continuous one-qubit rotation about the $X$, $Y$, or $Z$ axes of the Bloch sphere, and are a common choice of gate-set. We will refer to a particular decomposition of a rotation as an \textit{SK sequence}:

\begin{equation}
    R = U_{k}U_{k-1}...U_{1}U_{0} \hspace{5mm} U_{j} \in \{H,S,T\}
\end{equation}

where the length of the sequence $k$ can vary depending on the rotation $R_{Z}$ and approximation error. These sequences or decompositions can approximate any rotation $R(\hat{n}, \theta)$ about the axis $\hat{n}$ by an angle $\theta$. It is sufficient to synthesize continuous rotations about one axis, because rotations about any axis can be synthesized from the gate set~$\{H,S,T\}$ and continuous rotations around one axis. We focus on approximating the $R_{Z}(\theta)$ rotation.

The algorithm we use to generate our SK sequences is known as \textit{gridsynth} ~\cite{gridsynth}. It decomposes $R_{Z}(\theta)$ rotations into the $\{H,S,T\}$ gate set, minimizing the number of $T$-gates. Non-transversal gates are the most challenging to implement in QEC, and $T$-gates cannot be implemented transversally for many error correcting codes. This includes codes such as the surface code \cite{Bravyi_1998, Fowler_2012}, the Steane code \cite{Steane_1996}, the Shor code \cite{Shor_1995}, or 2D color codes \cite{Landahl_2011}, and known methods for implementing them non-transversally such as magic state distillation ~\cite{Knill_1996, Knill_1997, Campbell_2017b} or cultivation \cite{Gidney_2024} is costly. The \textit{gridsynth} algorithm is able to obtain solutions with $T$-counts $t + \mathcal{O}(\log(\log(1/\epsilon)))$, where $t$ is the $T$-count of the second most optimal solution. We refer the reader to \cite{gridsynth} for more details, but we do note a few aspects of interest.

\begin{figure}
    \centering
    \includegraphics[width=\textwidth]{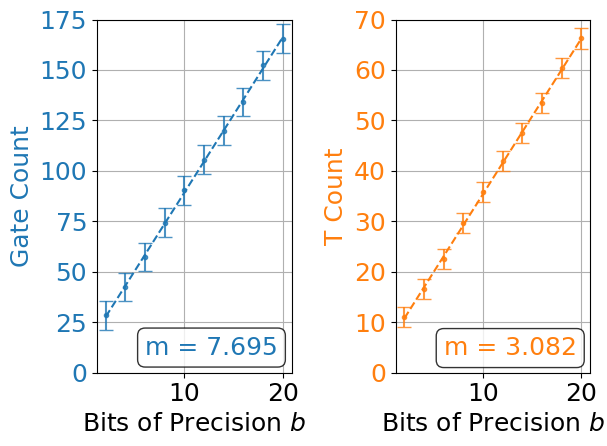}
    \caption{Average gate count (left) and average $T$-gate count (right) for sequences generated by gridsynth as a function of the bits of precision $b$, where $\epsilon = 2^{-b}$. Error bars are the standard deviation over an ensemble of $100$ sequences. Fitted slopes are $7.695$ for gate count and $3.082$ for $T$-gate count.}
    \label{fig:Gate count graph}
\end{figure}

The \textit{gridsynth} algorithm takes as input a random seed, the desired rotation angle $\theta$, the binary precision $\epsilon = 2^{-b}$ in realizing that rotation, and an ``effort'' integer to determine how much time the algorithm should spend finding an optimal solution. We may vary these parameters in order to generate multiple unique sequences for the same rotation. By varying the desired rotation angle $\theta$ by a value smaller than $\epsilon$, we can find multiple different sequences $R_{i}$ where $i=1,2,...,r$ that satisfy our precision constraint. These sequences may have slight variations in $T$-counts but are close enough to be grouped together in an equivalent ensemble. The gate count and $T$-count for these sequences increase linearly as a function of binary precision $b$, as can be seen in Figure \ref{fig:Gate count graph}.

We note that in order for a precision of $\epsilon = 2^{-b}$ to affect the results of an experiment implementing a single rotation, it is necessary to also have a sufficient sampling budget. This requires a number of samples:
\begin{equation}
    n \sim \mathcal{O}(\epsilon^{-2}) = 4^{b}
\end{equation}
which grows very quickly with the desired precision. This shows that for any fixed number of samples there is an implied maximum value $b_{\rm max}$ of $b$ beyond which further increases in sequence precision are not distinguishable from sequences obtained with precision $b_{\rm max}$.  Further discussion of sampling noise can be found at the end of Section \ref{subsec:Noiseless Simulation}.

\subsection{SKARC Algorithm}
\label{subsec:SKARC Algorithm}

We now outline our randomization procedure \textit{Solovay Kitaev and Randomized Compilation} (SKARC) for a given quantum circuit $C$ containing different $R_{Z}$ gates and a total sampling budget $N$.

For each instance of an $R_{Z}$ gate we wish to compile in $C$, we generate an ensemble of unique gate sequences $R_{i}, \hspace{2mm} i=1,2,...,r$ using \textit{gridsynth} as described in Section~\ref{sec:Sequence Ensemble}. These sequences can then be randomly selected with or without replacement in place of each corresponding $R_{Z}$ rotation in $C$. It is possible to randomize over all possible combinations of compiled $R_{Z}$ gates, but usually unnecessary to see the benefits of randomization, so oftentimes the total number of randomized circuits can be truncated, as will be explored in this work in Section \ref{subsec:Noiseless Simulation}. We focus on a circuit containing a single $R_{Z}$ gate, but our results are easily extended to circuits with multiple rotations. Once all circuits have been compiled this way, each randomly compiled circuit is measured $N/r$ times, dividing the total sampling budget between the compilations. The resulting $N$ measurement outcomes are then combined into one distribution for analysis, effectively averaging the measurement results over $r$ different decompositions of the $R_{Z}$ gate.

We study the performance of our procedure via state tomography. We create and execute circuits for each SK sequence $R_{i}$ and then measure each of these circuits in the $X$, $Y$, and $Z$ bases. This results in $3r$ different sets of measurement outcomes that can be used to reconstruct $r$ Bloch vectors $\vec{s}_{i}$, one for each decomposition:
\begin{equation}
    \vec{s}_{i} = \langle X \rangle \hat{i} + \langle Y \rangle \hat{j} + \langle Z \rangle \hat{k}
\end{equation}

An ensemble of $m$ such vectors determines an average vector. This can be done either by combining the measurement results from each sequence together and calculating a new Bloch vector as described above, or simply taking the average of the vectors obtained from each sequence. We define this mean vector for a particular choice of $m$ sequences indexed by $\{ i_{1}, i_{2}, ... i_{m} \}$ as:
\begin{equation}
    \vec{s}_{\{ i_{1}, i_{2}, ... i_{m} \}} = \frac{1}{m} \left( \vec{s}_{i_{1}} + \vec{s}_{i_{2}} + ... + \vec{s}_{i_{m}} \right)
\end{equation}

In order to quantify the performance of SKARC, we consider the trace distance between the target quantum state and the mean quantum state from averaging. As these decompositions are for single-qubit rotations, the trace distance $D(\vec{r}, \vec{s})$ is equivalent to half the Euclidean distance between the vectors on the Bloch sphere, where:
\begin{equation}
    D(\vec{r}, \vec{s}) = \frac{|\vec{r} - \vec{s}|}{2}
\end{equation}
A smaller trace distance $D(\vec{R_{Z}}, \vec{s}) < D(\vec{R_{Z}}, \vec{s_{i}})$ indicates that we have better realized the target rotation. Here $\vec{R_{Z}}$ is the vector corresponding to the target $R_Z$ rotation, $\vec{s}$ is the mean vector for the entire ensemble, and $\vec{s_{i}}$ is the vector for a single sequence chosen at random or optimized for its cost.

Generating an ensemble of $r$ sequences is the first step in employing averaging over sequences. We will use noiseless simulations to evaluate the effect of randomization on the accuracy of the results, as in~\cite{Hastings_2016, Campbell_2017, Seiseki_2024}. In simulations with coherent noise added, and in experiment, we are evaluating the effect of randomization as a form of randomized compilation to mitigate coherent noise.

Across the ensemble of $r$ total sequences generated by gridsynth, some of them are going to be more robust to certain types of noise than others. We can evaluate the variation between different ensembles of size $m<r$ by sampling subsets of our overall ensemble of size $m$. It is intractable to consider all possible combinations for $r = 100$ unique sequences, so instead we randomly choose at most $q=1000$ different ensembles of size $1\leq m \leq 100$ for simulation and $1\leq m\leq 20$ for experiment. We sample these sub-ensembles randomly with replacement and calculate the mean trace distance as:
\begin{equation}
    D(m) = \frac{1}{q} \sum_{j = 0}^{q} D(\vec{s}_{\{ i_{1}, i_{2}, ... i_{m} \}_{j}}, \vec{R_{Z}})
\end{equation}
where $\vec{R_{Z}}$ is the Bloch vector resulting from a noiseless, continuous $R_{Z}$ rotation.

This average trace distance across $1000$ possible combinations of sequences for a given ensemble size is then used to determine the efficacy of SKARC. In Section \ref{sec:Results}, we compare the average error with and without randomization as a function of the size of the ensemble that we average over. We should expect that as the size of the ensemble of distinct sequences increases, our error metrics will converge to some finite value. Moreover, by studying the characteristics of the ensemble of rotations implemented by our ensemble of circuits, we can learn more about the limits of SKARC.

\section{Coherent Noise Model}
\label{sec:Coherent Noise Model}

The coherent noise model we use in the remainder of this work is to multiplicatively over-rotate each rotation comprising a $H$ gate in our circuit by a fraction $1+\delta$, for some error strength $\delta$. This gate-dependent coherent noise model is similar to those used in \cite{Zhang_2022, Mueller_2023}, and the original RC study \cite{Wallman_2016}, though the latter uses an additive over-rotation error rather than a multiplicative one.  We only model simple, single-qubit coherent errors, as this work is focused only on single-qubit rotations. We only include $H$ gates because on the QSCOUT ion trap the $S$ and $T$ gates are implemented with virtual $R_Z$ rotations, and incur no error. The $H$ gates are implemented as a product of gates $X\sqrt{Y}$ and each of these gates is over rotated by $\delta$ so that $X(\delta)=R_x(\pi(1+\delta))$ and $\sqrt{Y(\delta)}=R_y(\frac{\pi}{2}(1+\delta))$.

We choose values of $\delta$ that vary between $[0, 10^{-2}]$. If $\delta$ were much larger than this, then it would be very difficult to resolve the desired $\epsilon$, even for a small number of bits of precision. On a real quantum device, the strength of this coherent over-rotation will depend on the error rate of the device in question and how much error correcting codes suppress errors.

The effect of this noise model on an individual SK sequence can be modeled as an erroneous single-qubit rotation. That is, if the noiseless approximate rotation is $R_{i}(\hat{a}, \theta)$, then the noisy rotation will be $\tilde{R}_{i} = R_{i} \mathcal{E}$. Where $\mathcal{E} = R(\hat{b}, \phi)$ is a rotation of the state away from the noiseless result about an axis $\hat{b}$ by an angle $\phi$. The axis and angle of this error will themselves be functions of the original axis $\hat{n}$, the original angle $\theta$, and the strength of the noise model $\delta$. However, the functional form of this error in terms of these parameters is non-trivial and sequence dependent. As a result, each sequence will experience this noise model in a unique manner. Averaging over many sequences will result in an ensemble of outputs converting the coherent noise to stochastic noise.

\section{Results}
\label{sec:Results}
\begin{figure*}[t]
    \includegraphics[width=0.95\columnwidth]{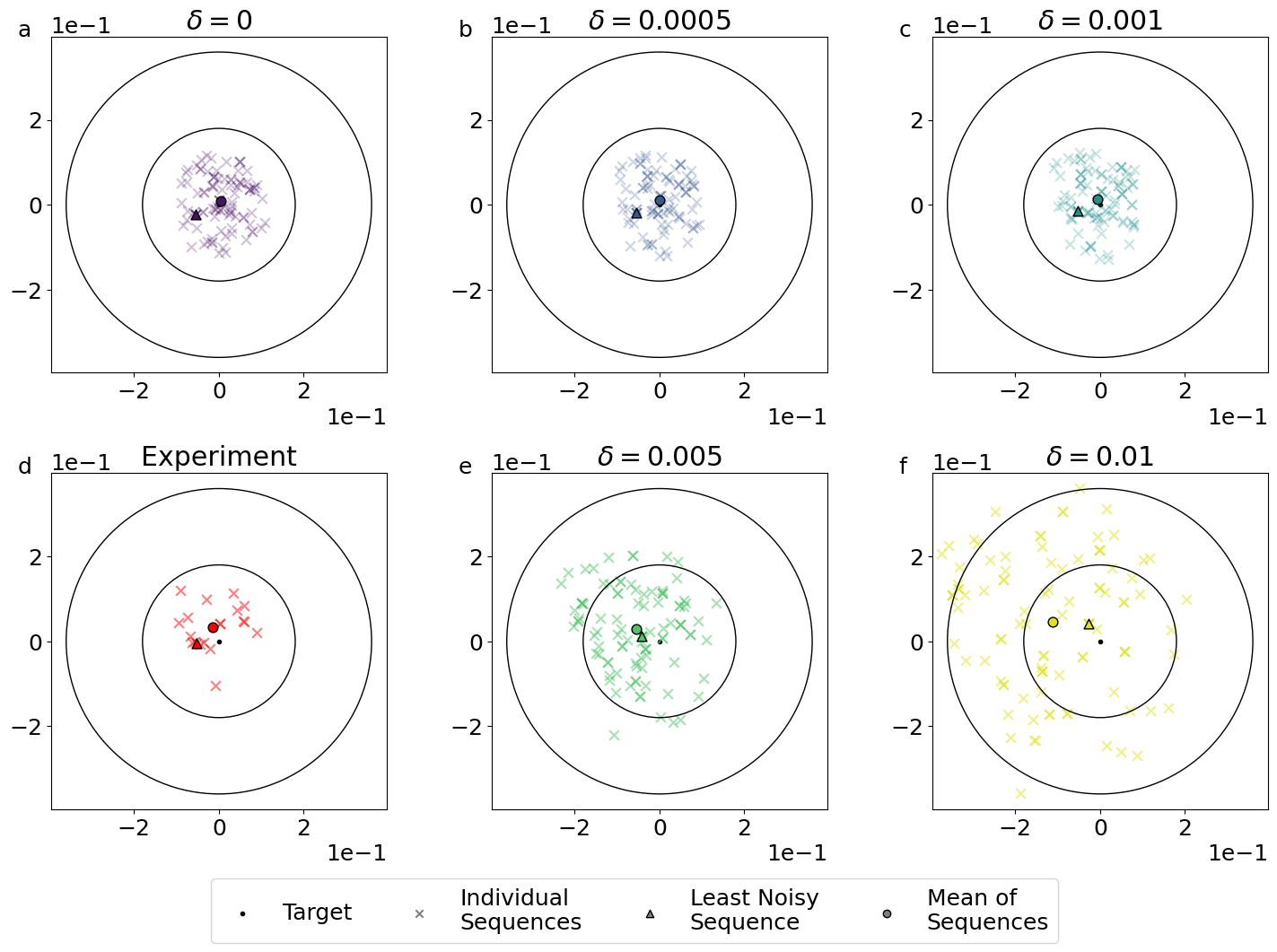}%
    \caption{Comparison of results from noiseless simulation, a coherent noise model, and experiment for precision $b=4$. Plots are ordered left to right, top to bottom based on trace distance of the mean Bloch vector to the target rotation. \textbf{(a, b, c, e, f)} 2D projections of Bloch sphere vectors as a function of the strength of the coherent noise model $\delta$. There is a bias in the vectors as a function of the noise, with larger $\delta$ shifting left of center due to the cumulative effects of the coherent noise manifesting as a single biasing coherent rotation. A negative $\delta$ would instead shift to the right. \textbf{(d)} Projection of experimental data is shown in red,with the mean Bloch vector having a similar trace distance compared to the sequence with the least theoretical error. Trace distance for the mean Bloch vector falls between $\delta=0.001$ and $\delta=0.005$. For coherent noise strengths $\delta \leq 0.001$ and experiment, randomization yields a lower trace distance than the sequence with the least theoretical error.}
    \label{fig:Large Target Graph b4}
\end{figure*}

\begin{figure*}[t]
    \centering
    \subfloat{
        \includegraphics[width=0.49\columnwidth]{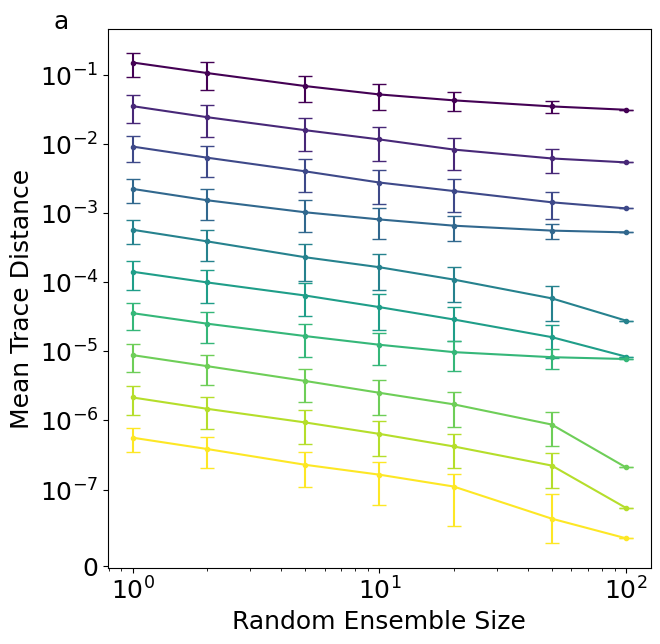}%
        \label{sfig:Noiseless Trace Distance}%
    }
    \hfill
    \subfloat{
        \includegraphics[width=0.49\columnwidth]{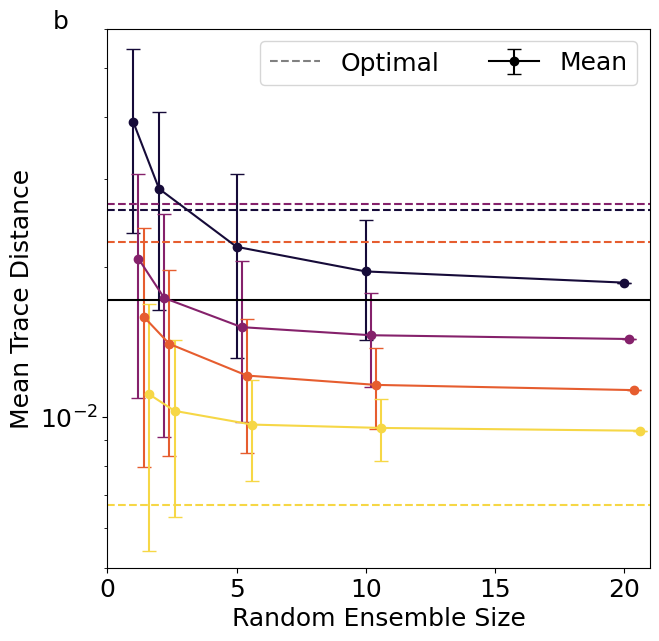}%
        \label{sfig:Experiment Trace Distance}%
    }
    \caption{\textbf{(a)} Simulation results of the mean trace distance across possible combinations of sequences for a given random ensemble size $m \leq 100$. Precision increases from $b=[2,20]$ in even increments from top to bottom. Only noiseless simulated data is shown. Accuracy increases with both precision and the size of the ensemble used for averaging. Error bars are taken as the standard deviation of the trace distances for all 1000 randomly chosen combinations of different sequences. \textbf{(b)} Experimental results of the mean trace distance. Precision increases from $b=[4,5,6,7]$ from top to bottom. Accuracy again increases with both precision and the size of the ensemble used for averaging. Error bars are calculated as for the simulated data, but for $m \leq 20$. Better accuracy can be obtained from choosing particular combinations of sequences as noted by the large error bars for smaller ensemble sizes. Trace distances for sequences with fewest $H$ gates are shown with dashed lines.}
    \label{fig:Ensemble Size Comparison}
\end{figure*}

\begin{figure*}
    \includegraphics[width=0.95\columnwidth]{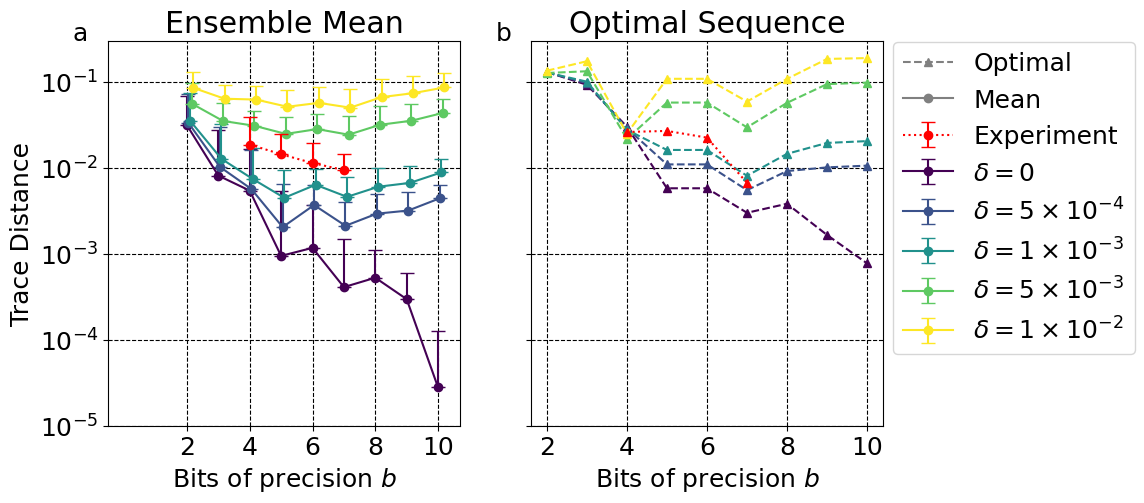}%
    \caption{Trace distance to the target rotation as a function of the bits of precision $b$. \textbf{(a)} Trace distance for the mean Bloch vector of the full ensemble of random sequences shown with solid lines and circles. The error bars are calculated from the variance in the components of the Bloch vectors across all sequences in the ensemble. Simulated data for different error strengths $\delta$ are shown with circles. Experimental results are shown with red dotted lines and circles. \textbf{(b)} Trace distance for the sequence in the ensemble with the fewest $H$ gates and least expected error shown with dashed lines and triangles. There are no error bars for this data, as there is only one sequence. Experimental results are shown with red dotted lines triangles.}
    \label{fig:Large Precision Comparison}
\end{figure*}

In this section, we present the results from two different wavefunction simulations using Sandia National Laboratories' JaqalPaq emulator \cite{Landahl_2020}, as well as an experiment on the QSCOUT device at Sandia National Laboratories \cite{Clark_2021}, a trapped-ion NISQ computer. Noiseless simulation results are discussed in Subsection \ref{subsec:Noiseless Simulation}. Results of simulations including the coherent noise model defined in Section~\ref{sec:Coherent Noise Model} are discussed in Subsection \ref{subsec:Coherent Simulation}. Experimental results are discussed in Subsection \ref{subsec:Experiment}. Each of these sections will reference and compare data shown in Figures \ref{fig:Large Target Graph b4}, \ref{fig:Ensemble Size Comparison}, and \ref{fig:Large Precision Comparison} which we will first describe here.

The simulated data comes from wavefunction simulations of the circuits that are equivalent to an infinite number of samples. In total, $r=100$ unique sequences were executed in simulation. The experimental data on the QSCOUT device was obtained from $24000$ samples per measurement basis, and executed $r=20$ sequences. The circuit is comprised of an $H$ gate to prepare the $|+\rangle$ state, followed by an $R_{Z}(\theta=1)$ rotation, followed by measurement. We choose the $|+\rangle$ state as our input state, as it is perpendicular to the axis of rotation on the Bloch sphere, and we chose the angle $\theta = 1$ as it is not a clean fraction of $\pi$, and therefore sufficiently far away from the unitaries generated by the gridsynth algorithm so as not to have a trivial decomposition even for many bits of precision.

To visualize the distribution of Bloch vectors originating from different sequences, we plot them on a 2D projection of the Bloch sphere. Figure \ref{fig:Large Target Graph b4} shows data for precision $b=4$. The Bloch vector corresponding to the target rotation $R_{Z}(\theta=1)$ is used as the normal vector for the 2D projection and is shown as a black dot in the center of the target. The statevectors for each of the $r$ SK sequences $\vec{s_{i}}$ are shown as colored crosses, with the mean of these statevectors $\vec{s}$ is shown as a circle. Experimental results are shown in red.

The triangle markers highlight the SK sequence with the smallest number of $H$ gates, which is the least affected by noise in our noise model. For all precisions tested except $b=7$ and $b=10$, the sequence with the fewest $H$ gates is also the sequence with the fewest $T$ gates, which is the normal output of the \textit{gridsynth} algorithm and the sequence with the least theoretical error for many error correcting codes. For precisions $b=7$ and $b=10$, the sequence with the least $H$ gates has the second-fewest $T$ gates instead.We note that the Ross-Selinger algorithm does not attempt to minimize the number of $H$ gates used in its compilation, so exact comparisons between experimental NISQ results for the least noisy sequence and simulated EFTQC results for the least noisy sequence are imperfect, even if they are identical amongst generated sequences.

Figure \ref{fig:Ensemble Size Comparison} shows simulated and experimental trace distances as a function of the size of the ensemble of sequences $m \leq r$ that we average over. The combinations of sequences are drawn randomly, with total number of possible combinations of sequences capped at $1000$. For the simulated data, the precision of the sequences ranges from $b=[2, 20]$ in even increments from top to bottom. For the experimental data, the precision of the sequences ranges from $b=[4,5,6,7]$ from top to bottom for the solid lines. The dashed horizontal lines in the experimental data represent the trace distance for the sequence with the fewest $H$ gates and least expected error, colored in the same order as the mean trace distance data.

Figure \ref{fig:Large Precision Comparison} compares trace distances to the target rotation with and without SKARC for simulated and experimental data as a function of the bits of precision $b$. The solid lines with correspond to trace distances for mean Bloch vectors over the entire ensemble. The dashed lines correspond to trace distances for the SK sequence with the fewest $H$ gates, and thus least error. Here the error bars are calculated as the variance in the each of the $X$, $Y$, and $Z$ components of the Bloch vectors across the ensemble of sequences. This error is then propagated through to the trace distance calculation. 

\subsection{Noiseless Simulation}
\label{subsec:Noiseless Simulation}
We first consider the noiseless case of $\delta = 0$. Figure \ref{fig:Large Target Graph b4}a shows the distribution of Bloch vectors realized by the ensemble of SK sequences. The mean of the ensemble lies close to the target rotation in the center. This is the expected result, as the approximation error inherent to each individual sequence is randomly distributed, and thus will cancel out in aggregate. Moreover, this mean lies closer to the center than the SK sequence output by gridsynth with the least theoretical error highlighted with a triangle. This shows that use of randomization over SK sequences can improve the accuracy of the aggregate result.

The average trace distance as a function of the ensemble size $m$ for a larger set of noiseless data for $b=[2,20]$ is shown in Figure \ref{sfig:Noiseless Trace Distance}. We see that the mean trace distance decreases as the ensemble size increases, indicating that randomizing over more SK sequences leads to quantum states that are closer to those of the target $R_{Z}$ rotation. We note that taking into account the lowest value of the error bars in Figure \ref{sfig:Noiseless Trace Distance}, the smallest possible trace distance would be achieved by averaging over less than the full circuit ensemble, which is expected. There will generally exist some particular combination of sequences that when averaged together will achieve a trace distance that is less than the trace distance for the average of the full ensemble. These sorts of algorithms are the basis of other probabilistic gate synthesis methods \cite{Campbell_2017, Seiseki_2024}. That is, deliberately picking two sequences that are equally and oppositely spaced from the true rotation will give a very accurate result, but it requires more work to implement this sort of weighting of different unitaries.

The noiseless results in Figures \ref{fig:Large Precision Comparison}a and b indicate that randomization improves accuracy. The trace distance for the mean Bloch vectors are about $\simeq 0.15$ of the trace distances for the optimal Bloch vectors averaged over varying precisions. This improvement is consistent as the precision is increased, and is similar to increasing the bits of precision by $2-3$. Using Figure \ref{fig:Gate count graph}, we see this is equivalent to a reduction of $\sim 5-10$ in the required number of $T$-gates needed to achieve the desired precision, and a reduction in total gate count of $\sim 15-25$.

We also examined the relationship between the number of samples used in simulation and the accuracy of the rotation. For an approximation error of $\epsilon=2^{-b}$, we would expect to need $N=\epsilon^{-2}=4^{b}$ samples to match. Using fewer samples than this means that imprecision due to sampling noise will overshadow any gains made by using more precise SK sequences. Using more samples than this means the opposite: that imprecision due to the approximation error of the sequence will overshadow any gains made by using more samples.

\begin{figure}[t]
    \centering
    \subfloat{\includegraphics[width=\textwidth]{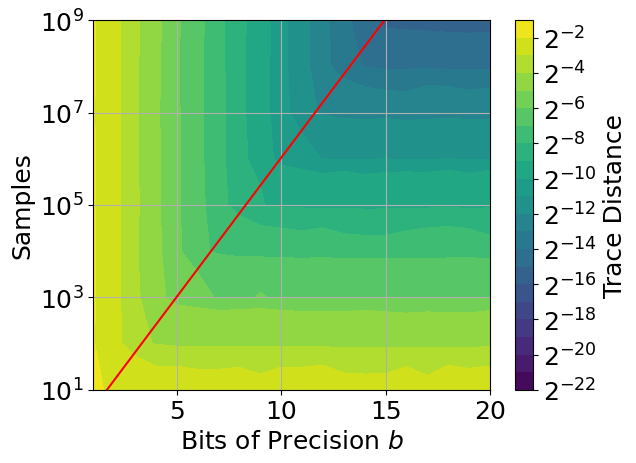}%
    \label{sfig:Sampling Contour Plot}}
    \\
    \subfloat{\includegraphics[width=\textwidth]{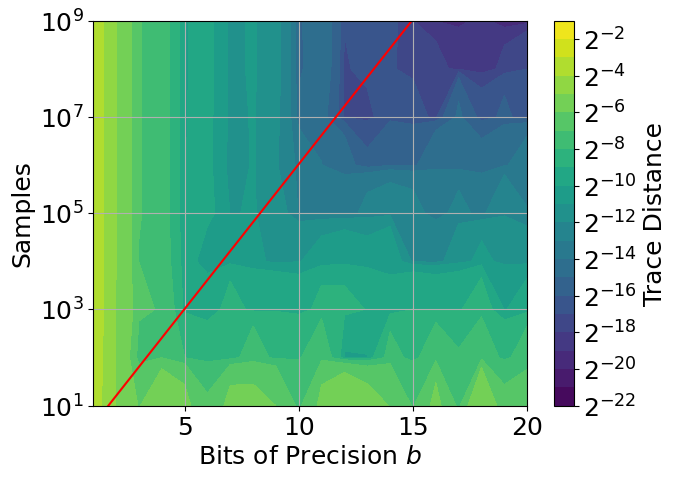}%
    \label{sfig:SKARC Sampling Contour Plot}}
    \caption{Trace distance to an ideal $R_{Z}$ rotation as a function of the number of samples and bits of precision. Results are shown without averaging over random sequences (top) and with randomization over the maximum ensemble size (bottom). The red line indicates the number of samples needed to match the approximation error $\epsilon = 2^{-b}$ of an SK sequence. Using more samples above this line leads to diminishing returns in accuracy. Using a higher precision sequence with fewer samples to the right of this line also leads to diminishing returns.}
    \label{fig:Sampling Contour Comparison}
\end{figure}

We see the necessity of both a large number of samples and many bits of precision in Figures \ref{sfig:Sampling Contour Plot} and \ref{sfig:SKARC Sampling Contour Plot}. The red line on these contour plots indicates the $N=4^{b}$ number of samples required for the theoretical best precision. We see that this lines up nicely with the shape of the contour itself. Improvements in precision (i.e. lower trace distance) can only be achieved by both increasing the number of samples and the precision of the sequence. This is also true for the randomized case. However, the randomized data generally show smaller trace distances than the non-randomized data, as expected. Using randomization, we can achieve better precisions without needing to increase the number of samples.

\subsection{Simulation with Coherent Noise}
\label{subsec:Coherent Simulation}
The question remains, do the improvements we see in the noiseless case hold in the presence of coherent noise? Single-qubit coherent errors manifest as unitary rotations on the Bloch sphere, and so in practice should be corrected in a manner similar to the synthesis error seen in the previous section. In order to answer this question, we repeat our analysis with the simple model of coherent noise described in Section \ref{sec:Coherent Noise Model}.

Figures \ref{fig:Large Target Graph b4}b, c, e, and f, show that each individual sequence experiences the coherent error differently, as evidenced by the varied distributions of sequence vectors $\vec{s_{i}}$ as a function of the coherent noise parameter $\delta$. However, the change is not totally random. The stronger the error strength, the more the ensemble shifts to the left, manifesting as a single coherent rotation on the surface of the Bloch sphere. A negative value of $\delta$ instead shifts the ensemble to the right.

The performance of SKARC in the presence of coherent error is shown in Figure~\ref{fig:Large Precision Comparison}. The trace distance to the target rotation for simulation data falls with increasing bits of precision for nonzero $\delta$ between $b=2$ and $b=7$, before rising again for nonzero coherent noise. This increase shows that noise has a limiting effect on precision. Too much noise and it is impossible to realize highly precise rotations, regardless of whether randomization is used.

This increase in error happens for two reasons. Primarily because the coherent error model shifts the overall distribution away from the target rotation. This effect compounds with longer sequences as they have more coherent error overall; hence, the error increases for more bits of precision. This can be seen for both the ensemble mean and the sequence with the least theoretical error. The second reason is that, even when the coherent noise is tailored by randomization, it has an effect as depolarizing noise. The mean vector of the ensemble has a magnitude slightly less than that of any individual sequence, and for large enough error strengths this will additionally increase error as a function of sequence length. For smaller coherent errors $\delta \sim 1 \times 10^{-4}$, this increase in trace distance is delayed until around $b=9$ bits of precision.

Comparing the simulated data in Figure \ref{fig:Large Precision Comparison}a and b, the trace distances for the mean Bloch vector are smaller than the trace distance of the sequence with the fewest $H$ gates and least expected error. The exceptions are for precision $b=4$ and for some coherent noise strengths $b=7$. We interpret this as the variability of the performance of the optimal sequence being larger than that of the mean, and therefore the optimal sequence sometimes outperforms the average, particularly when taking the overall shift of the ensemble into account.

\subsection{Experimental Results}
\label{subsec:Experiment}

We performed experiments on the QSCOUT device at Sandia National Laboratories \cite{Clark_2021}, a trapped-ion NISQ computer. As the experiments are performed on a single ion and the depths of the circuits are considerable, driving the qubit transition of 12.6428 MHz directly with globally-acting microwaves (with seconds-long coherence) is preferable over laser-based Raman gates. We do so with an external microwave horn in which the microwaves are controlled by custom control hardware system that allows for arbitrary phase, amplitude, and duration control \cite{Clark_2021}. We used $N=24000 \sim 2^{14} \sim 4^7$ samples for each basis measurement due to time constraints, corresponding to a sampling error of $\epsilon = 1/\sqrt{N} \sim 0.006$. This means that for higher precisions than $b=7$, finite sampling error dominates as opposed to the approximation error, as was discussed in Section \ref{subsec:Noiseless Simulation}. As such, we restrict our experimental data to the range $b=[4,5,6,7]$.

The Bloch vectors for each of the sequences $R_{i}$ are shown in Figure \ref{fig:Large Target Graph b4}d for $b=4$ as red crosses, with the mean shown as a red circle. For precision $b=4$, the spread of the experimental data is consistent with a value of $\delta$ between $1\times10^{-3}$ and $\delta = 5\times10^{-3}$ in simulations. The Bloch vector corresponding to the sequence with the fewest $H$ gates and least expected error is shown with a red triangle as part of the ensemble.

Figure \ref{sfig:Experiment Trace Distance} shows the trace distance of the mean of different ensemble sizes $m$ across different bits of precision $b$. The trace distance decreases with higher precision and as the size of the random ensemble increases. The error bars in Figure \ref{sfig:Experiment Trace Distance}  are calculated as the standard error on the mean of the distribution of trace distances for each possible combination of sequences for a given ensemble size. For most precisions, the sequence with the fewest $H$ gates and least theoretical error is further away than the mean sequence from averaging over the whole ensemble (the data points on the right). However, this is not the case for precision $b=7$, where this shorter sequence is much more accurate than the average. 

In Figure \ref{fig:Large Precision Comparison}b we give the average trace distance for both simulations including coherent noise and for experiment. In Figure \ref{fig:Large Precision Comparison}b, we compare the trace distance of the sequence with the fewest $H$ gates. Again we find that the experimental data corresponds to an error rate of $\delta \sim 2\times10^{-3}$. We see larger trace distances for the sequence with the fewest $H$ gates when compared to averaging across different precision, though the experimental data point at $b=7$ is lower than expected, as was discussed previously. Moreover, this error rate is in agreement with, if slightly larger than the estimated error in single-qubit gates from other experimental benchmarks of the QSCOUT device \cite{Clark_2021}.

We see a reduction in the trace distance using SKARC to compute the mean vector compared to the sequence with the fewest $H$ gates, except for the case of $b=7$. Even in the presence of sampling noise and NISQ-era device noise, averaging the Bloch vectors of the ensemble renders a result that is closer to the target rotation, with improvement similar to that seen for the coherent noise model. 

\section{Conclusions}
\label{sec:Conclusions}

We find that it is possible to reduce the effects of a wide variety of physical errors and approximation errors inherent to Solovay-Kitaev decompositions of continuous rotations by averaging over an ensemble of distinct gate sequences. Multiple similar-yet-distinct sequences are generated using the \textit{gridsynth} algorithm by varying the target rotation as described in Section \ref{subsec:Generating Sequences}. When acting on a single-qubit quantum state, these sequences result in an ensemble of Bloch vectors that are clustered around the target rotation state as discussed in Section \ref{subsec:SKARC Algorithm}. By averaging the results of multiple circuits corresponding to each sequence, we are able to obtain quantum states that are closer to the ideal rotation as measured by the trace distance between the states.

We presented a noiseless simulation in Section \ref{subsec:Noiseless Simulation} and analyze the ensemble of sequences that results. Across multiple desired approximation precisions $\epsilon=2^{b}$ we find that the average of the full ensemble of sequences yields a Bloch vector that outperforms the sequence with the fewest $H$ gates and least expected error. These results mirror those of other more sophisticated probabilistic gate synthesis algorithms such as convex hull finding \cite{Campbell_2017, Seiseki_2024}. We study the effects of sampling noise on our method, and find the expected trend: The number of samples needed scales exponentially with the desired precision of the sequence, and under-sampling the circuit will lead to errors regardless of randomization. Similarly, over-sampling the circuit will be useless if the precision of the rotation is not increased in turn.

Improvement using randomization is maintained under a coherent noise model. In Section \ref{subsec:Coherent Simulation} we explore the effects of a simple multiplicative coherent noise model on the rotations comprising $H$ gates in our circuit, mirroring performance on QSCOUT, a NISQ trapped-ion computer. Under such a model, SKARC can mitigate coherent errors and reduce the trace distance to the target rotation, though a coherent bias remains and limits the effectiveness of the method at realizing precise rotations with large coherent error. For EFTQC architectures with small code sizes or where error correction is not performed after every single qubit gate, this technique may improve accuracy and/or reduce costs.

Lastly, we study the technique on the QSCOUT device in Section \ref{subsec:Experiment}. We find that randomization can improve the accuracy of rotations on the device, though as with the coherent noise model, the resulting mean Bloch vector is still biased. The device performance is most closely replicated by a coherent error strength of $\delta \sim 2\times10^{-3}$, or a $0.2\%$ over-rotation, which matches with experimental benchmarks. Additionally, the number of samples required to realize high precision remains an obstacle in the near-term.

These results provide a NISQ baseline for single-qubit gate synthesis algorithms which we can compare against EFTQC devices in the future. Further research should be carried out for a wider variety of noise models that better capture the effects of error-correcting codes. Additionally, it would be of interest to benchmark all or a portion of this technique on an error-corrected qubit to see if the results hold under that paradigm. Methods such as SKARC that rest at the intersection of error mitigation and error correction will become more important as the EFTQC era approaches.

\section{Acknowledgments}
This project was funded by the U.S. Department of Energy, Office of Science, Office of Advanced Scientific Computing Research Quantum Testbed Program. Sandia National Laboratories is a multi-mission laboratory managed and operated by National Technology and Engineering Solutions of Sandia, LLC, a wholly owned subsidiary of Honeywell International Inc., for the U.S. Department of Energy’s National Nuclear Security Administration under contract DE-NA0003525.  This paper describes objective technical results and analysis. Any subjective views or opinions that might be expressed in the paper do not necessarily represent the views of the U.S. Department of Energy or the United States Government.  SAND2025-03150O.

\bibliographystyle{unsrt} 
\bibliography{citations}

\end{document}